# First-principles study on the superconductivity of N-doped $fcc$-LuH$_3$

Zihao Huo[1], Defang Duan[1,*], Tiancheng Ma[1], Qiwen Jiang[1], Zihan Zhang[1], Decheng An[1], Fubo Tian[1], Tian Cui[1,2, *]

[1]*State Key Laboratory of Superhard Materials, College of Physics, Jilin University, Changchun 130012, China*
[2]*Institute of High Pressure Physics, School of Physical Science and Technology, Ningbo University, Ningbo 315211, China*
*Corresponding author: duandf@jlu.edu.cn, cuitian@nbu.edu.cn

**Abstract:** Recently, room-temperature superconductor has been claimed in a nitrogen-doped lutetium hydride at near-ambient pressure [Nature 615, 244 (2023)]. Using X-ray diffraction (XRD) and Raman spectra analysis, the authors believed that the superconducting properties can most probably be attributed to $Fm\bar{3}m$-LuH$_{3-\delta}$N$_\varepsilon$. Here, we systematic study the phase diagram of Lu-N-H at 1 GPa by using first-principle theory and find that there have no thermodynamically stable ternary compounds. Besides, we analyzed the dynamically stability and superconducting properties of N-doped $Fm\bar{3}m$-LuH$_3$ using virtual crystal approximation (VCA) and supercell method. Our theoretical results show that the $T_c$ of N-doped LuH$_3$ cannot reach the level of room-temperature.

## I. Introduction

The research of superconductivity in hydrides has received intensively attentions since the high critical transition temperature $T_c$ almost approaches room temperature[1-3]. In 2014, hydrogen sulfide H$_3$S was predicted to be a high-temperature superconductor with a $T_c$ of 191-204 K[4], which was confirmed by the experimentally measured $T_c$ of 203 K at 155 GPa[5, 6]. Following this success, several new hydrides were predicted and synthesized[7-15]. Recently, Hao *et al.* systematic study the Lu-H system and found that filled *f*-shells would lead a strong electron-phonon interaction. They predicted that $Im\bar{3}m$-LuH$_6$ has a $T_c$ of 273 K at 100 GPa[13]. Then extensive research has been conducted on the superconducting property of lutetium hydrides under pressure. $Fm\bar{3}m$-LuH$_3$ have been successfully synthesized with $T_c$ of 12 K at 122 GPa[16]. Moreover, $Pm\bar{3}n$-Lu$_4$H$_{23}$ have been obtained with $T_c$ of 71 K at 218 GPa[17]. Especially, a very recent experimental study has reported superconductivity in the Lu-N-H system at near-ambient pressure (about 1 GPa), with the highest $T_c$ of 294 K[18]. This work has ignited people's dream of achieving room-temperature superconductivity near ambient pressure. The authors believed that the room-temperature superconductor in the Lu-N-H system is most probably be attributed to $Fm\bar{3}m$-LuH$_{3-\delta}$N$_\varepsilon$, which can be seen as the nitrogen atoms doped into $Fm\bar{3}m$-LuH$_3$.

Although this report set a landmark in scientific community, there are still many open questions surrounding this important discovery. For instance, the exact stoichiometry of hydrogen and nitrogen, and their respective atomistic positions are still elusive. Moreover, debate soars within the scientific community as a group of scientists are skeptical about the results, and extra experimental confirmations. A following experimental observation indicate that the pressure-induced color change of LuH$_2$ is similar to that of N-doped lutetium hydride[19]. At the same time, a similar compound LuH$_{2\pm x}$N$_y$ have been obtained from Xue *et al.* [20], but no superconductivity has been obtained at pressure range from 1 to 6 GPa. Hence, to clarify the mechanism of superconductivity properties, it is highly desirable to know the crystalline structure from the theoretical side.

In this paper, we performed a comprehensive first-principles study on the Lu-N-H system at 1 GPa. However, no thermodynamically stable compounds in Lu-N-H system have been found at this pressure. Then we analyzed the dynamically stability and superconducting properties of N-doped $Fm\bar{3}m$-LuH$_3$ by using VCA and supercell method. We found that the introduction of nitrogen atoms into $Fm\bar{3}m$-LuH$_3$ may slightly enhance its $T_c$, but it would decrease its dynamics stability and their $T_c$ cannot reach the level of room-temperature. Our theoretical results show that the $T_c$ of N-doped LuH$_3$ cannot reach the level of room-temperature.

**II. Computational details**

At pressure of 1 GPa, we performed the variable-composition crystal structure searches in Lu-N-H system with approximately 10000 structures using *ab initio* Random Structure Searching[21] (AIRSS) code. Then we re-optimized the structures using the *ab initio* calculation of the Cambridge Serial Total Energy Package[22] (CASTEP). The on-the-fly ultrasoft pseudopotentials with the valence electrons $1s^1$ for H, $2s^2 2p^3$ for N, and $4f^{14} 5s^2 5p^6 5d^1 6s^2$ for Lu were employed with a kinetic cutoff energy of 800 eV. The Brillouin zone was sampled with a *k*-point mesh of $2\pi \times 0.03$ Å$^{-1}$ to make the enthalpy calculations well converged to less than 1 meV/atom. The structural relaxations were carried out using the projector-augmented wave[23, 24] (PAW) potentials, as implemented in the Vienna *ab initio* simulation packages[25] (VASP) with the energy cutoff of 600 eV. The exchange-correlation functional was described using Perdew-Burke-Ernzerh (PBE) of generalized gradient approximation[26] (GGA).

We investigated the pressure dependence of the superconductivity of $Fm\bar{3}m$-LuH$_3$ at 0.5-3% doping of N by using VCA method. Electronic structure calculations, phonon, and electron-phonon coupling (EPC) were calculated with the QUANTUM ESPRESSO[27] (QE) package. The PAW pseudopotentials with the valence electrons $1s^1$ for H, $2s^2 2p^3$ for N, and $5s^2 5p^6 5d^1 6s^2$ for Lu were used in QE package. Self-consistent electron density was evaluated by employing *k*-mesh of 20×20×20. Phonon and EPC were calculated using *q*-mesh of 5×5×5. The superconducting transition temperatures are estimated through the Allen-Dynes-modified McMillan equation[28] (A-D-M) with correction factors and the Coulomb pseudopotential[29] $\mu^* = 0.10$ and 0.13.

Then we prepared a conventional cell of $Fm\bar{3}m$-LuH$_3$, including 4 formula units (f.u.). The calculations were performed for a range of pseudostoichiometries Lu$_4$H$_m$N$_{12-m}$: Lu$_4$H$_{11}$N, Lu$_2$H$_5$N, Lu$_4$H$_9$N$_3$, and LuH$_2$N, corresponding to nitrogen-to-hydrogen atomic ratios 0.09, 0.2, 0.33, and 0.5. Next we calculated the formation enthalpy, electronic structure, and superconductivity of these structures at 1 GPa, 10 GPa, and 50 GPa. The formation enthalpy of Lu$_4$H$_m$N$_{12-m}$ has been calculated by the following equation:

$$H^f = H(\text{Lu}_4\text{H}_m\text{N}_{12-m}) - 4H(\text{LuH}_3) - mH(\text{N}) + \frac{m}{2}H(\text{H}_2) \qquad (1)$$

## III. Results and discussion

We performed a random structure search for Lu-N-H system and constructed the ternary phase diagram (convex hull) at 1 GPa, as shown in Fig. 1(a). Part of the binary compounds were adopted from previous paper[30-33]. We find that there has no ternary compound lie on the convex hull at 1 GPa. So there is no ternary Lu-N-H compound that can maintain thermodynamically stable at this pressure. Besides, the $P\bar{3}m1$-Lu$_2$H$_2$N have been found with the enthalpy is ~3 meV/atom above the convex hull, the detailed information of structure parameters have been listed in Table S1 of the Supplemental Material. According to the inorganic crystal structure database, 20% of experimentally synthesized materials are metastable[34, 35]. So we calculated the XRD data of this compound. Fig. 1(b) shows a comparison of XRD data based on previous experiments and our calculations. We can find the simulated XRD of $P\bar{3}m1$-Lu$_2$H$_2$N obviously deviates the experimental one, excluding its existence in the N-doped lutetium hydride superconductor. Therefore, we did not find any suitable ternary compound candidate structures to explain the experimental observations.

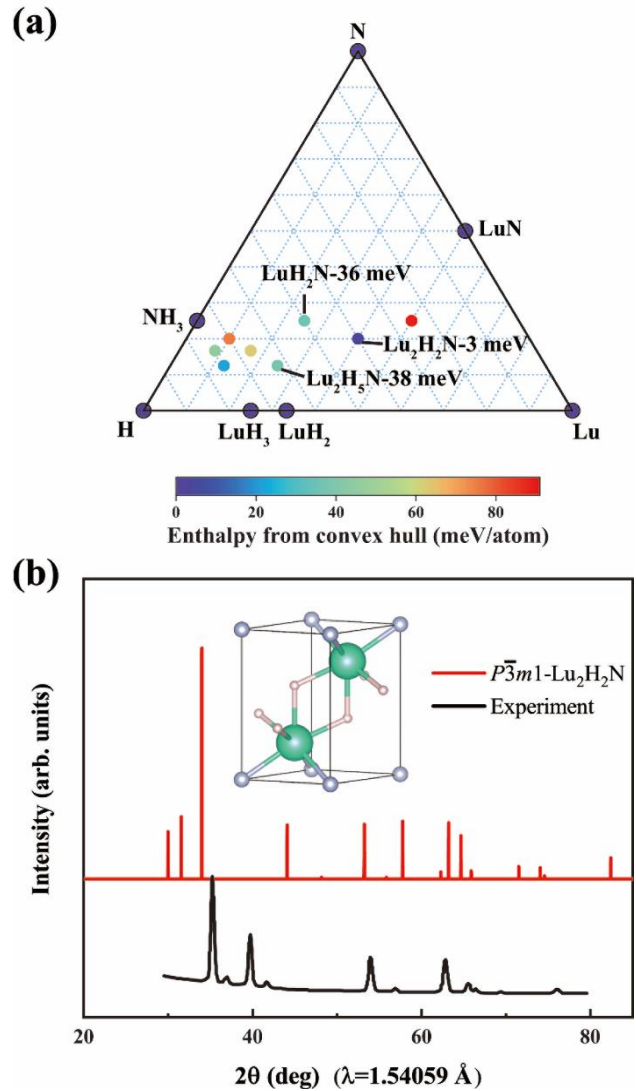

Fig. 1. (a) The ternary phase diagram (convex hull) of the Lu-N-H system at 1 GPa. Colored square means the value of enthalpy form convex hull. Squares with black solid boundary denote stable phase. (b) Comparison of XRD of reported experimental data and our calculated date at ambient pressure. Upper red curve represents the simulated XRD pattern from this work, and bottom black curve comes from experimental data[18]. The inset exhibits the crystal structure of $P\bar{3}m1$-$Lu_2H_2N$. The green, gray, and pink spheres depict Lu, N, and H atoms, respectively.

Next we investigated the doping effect on N-doped $LuH_3$ by using the supercell method. We constructed the model of $Lu_4H_mN_{12-m}$ by replacing hydrogen atoms with nitrogen atoms, where m = 8-11. And for each concentration, the total energy was calculated as each new configuration replaced $O$ with $T$ occupancy, one atom at a time, until all $O$ sites were converted to $T$ sites. The formation enthalpies of $Lu_4H_mN_{12-m}$ have been calculated at 1 GPa, 10 GPa and 50 GPa. Fig. 2 shows the results of formation enthalpy of $Lu_4H_mN_{12-m}$ at different pressure. At 1 GPa, we found that only the enthalpy of formation of $Lu_4H_{11}N$ is negative (about -24 meV/atom), which means this compound probably be thermodynamically stable. A similar situation also occurred at 10 GPa (about -32 meV/atom for $Lu_4H_{11}N$). It is indicated that only $Lu_4H_{11}N$ can form within 10 GPa. When the pressure reaches 50 GPa, $Lu_4H_{11}N$, $Lu_2H_5N$, $Lu_4H_9N_3$, and $LuH_2N$ can be formed. So only $Lu_4H_mN_{12-m}$ with lower nitrogen doping can be formed at low pressure condition. However, as the pressure increases, the nitrogen doping concentration that can be incorporated also increases. Besides, it is apparent that there is a preference for one nitrogen atom to occupy a $T$ site in $Lu_4H_mN_{12-m}$ at 50 GPa. And in the whole pressure range we studied, nitrogen atom preferentially occupies $T$ site in $Lu_4H_{11}N$.

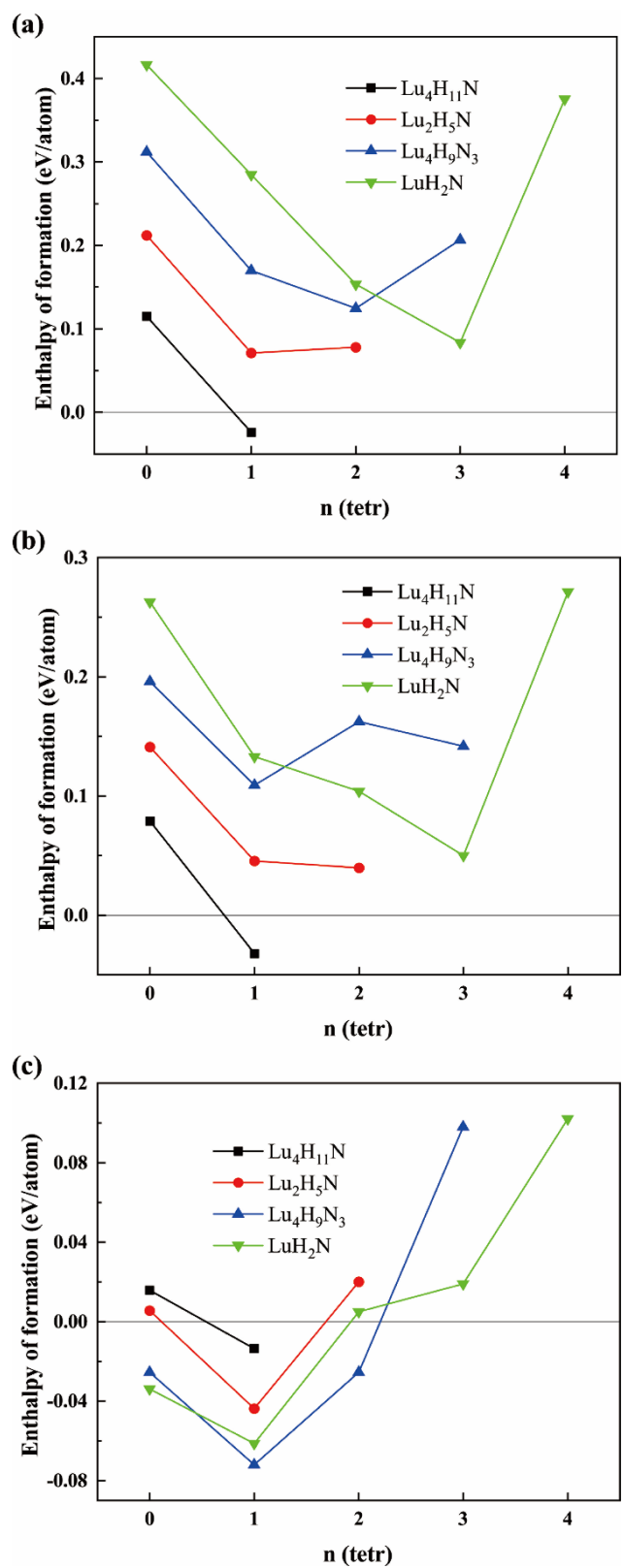

Fig. 2. Calculated enthalpy of formation of $Lu_4H_mN_{12-m}$ as a function of $T$-site occupancy at (a) 1 GPa, (b) 10 GPa, and (c) 50 GPa, obtained by starting with pure $O$ occupancy and progressively moving N to $T$ sites. n(tetr) is the number of nitrogen atoms out of the total in each configuration occupying $T$ sites.

Then we calculated the dynamically stability and the superconducting properties of $Lu_4H_mN_{12-m}$. Their $T_c$s are estimated through A-D-M with correction factors (Table S2 of Supplemental Material). At 1 GPa and 10 GPa, we find that the $Lu_4H_{11}N$ cannot be dynamically stable. In order to investigate this phenomenon, we computed the phonon spectrum of $Fm\bar{3}m$-$LuH_3$ at 10 GPa. We also calculated the enthalpies of $Fm\bar{3}m$-$LuH_3$ and $P\bar{3}c1$-$LuH_3$, which have been synthesized at ambient pressure[31]. The results have been plotted at Fig. S1 of Supplemental Material. At pressure below 10 GPa, the enthalpy of $P\bar{3}c1$-$LuH_3$ is lower than that of $Fm\bar{3}m$-$LuH_3$, indicating that $P\bar{3}c1$-$LuH_3$ is a thermodynamically more stable phase (see Fig. S1(a)). Besides, the existence of imaginary frequency in the whole Brillouin zone indicated the dynamically unstable of $Fm\bar{3}m$-$LuH_3$ at 10 GPa. So the $Fm\bar{3}m$-$LuH_3$ cannot be thermodynamically stable and dynamically stable at pressure below 10 GPa. And the dynamically unstable of $Lu_4H_{11}N$ may caused by the instability of $Fm\bar{3}m$-$LuH_3$ under pressure. At 50 GPa, only the $Lu_2H_5N$ can be dynamically stable. The crystal structure, phonon spectrum, and electron-phonon coupling property of $Lu_2H_5N$ are illustrated in Fig. 3. $Lu_2H_5N$ ($Lu_4H_{10}N_2$) can be considered as composed of two nitrogen atoms occupying the $T$ site and $O$ site of $Fm\bar{3}m$-$LuH_3$. And the introduction of nitrogen atoms would cause lattice mismatch in $Fm\bar{3}m$-$LuH_3$ and resulting in $R3m$ space group. The phonon and EPC calculation shows that the contribution of mid-low frequency (13~25 THz) phonons to the EPC is the highest (about 46% to total $\lambda$), while high frequency (45-55 THz) phonons hardly contribute to the EPC. It is indicated that the EPC of $R3m$-$Lu_2H_5N$ is mostly contributed by the mid-low frequency phonons. The calculated EPC parameter $\lambda$ for $R3m$-$Lu_2H_5N$ at 50 GPa is 0.82. Using the calculated logarithmic average frequency $\omega_{\log}$, along with the Coulomb pseudopotential $\mu^*$ value of 0.1, the resultant $T_c$ value is 27 K.

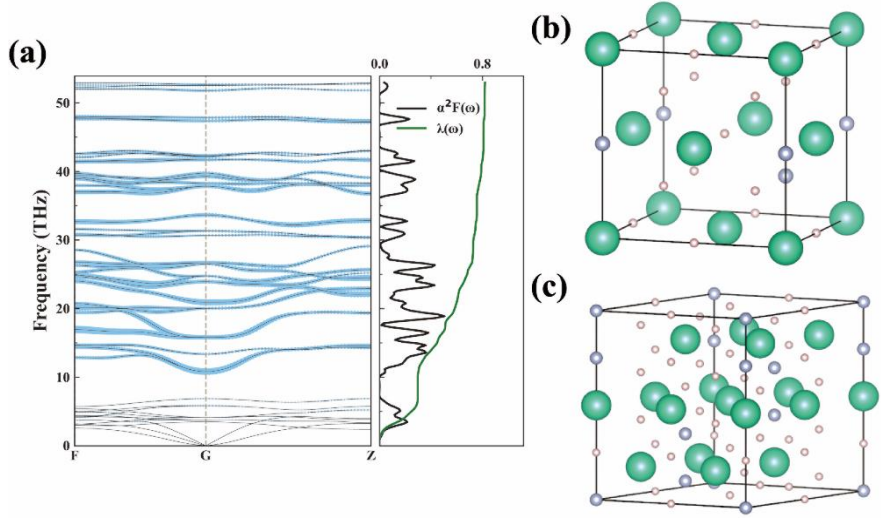

Fig. 3. (a) Phonon spectra, Eliashberg phonon spectral function $\alpha^2F(\omega)$, and the electron-phonon integral $\lambda$ for $R3m$-$Lu_2H_5N$ at 50 GPa. The magnitude of the phonon linewidths is illustrated by the size of the blue circles. (b-c) The crystal structure of $R3m$-$Lu_2H_5N$ in the framework of $Fm\bar{3}m$-$LuH_3$ and conventional cell, respectively. The green, gray, and pink spheres depict Lu, N, and H atoms.

In the case of lower nitrogen doping concentrations, we employed the VCA method to investigate the pressure dependence of the superconductivity in N-doped $Fm\bar{3}m$-$LuH_3$. Fig. 4(a) shows the N-doped concentration dependence of the minimum dynamically stable pressure. We find that the N-doped concentration are closely related to the minimum dynamically stable pressure. The minimum dynamically stable pressure of N-doped $Fm\bar{3}m$-$LuH_3$ increases from 25 GPa to 70 GPa when the doping concentration increases from 0% to 2%. Besides, at a doping concentration of 3%, the structure cannot maintain stability within the pressure range we studied. Therefore, the introduction of N atoms would decrease the structural stability of $LuH_3$. Besides, we calculated the $T_c$ of N-doped $LuH_3$ with doping concentrations ranging from 0 to 1.5% at 50 GPa (see Fig. 4(b)) to facilitate the investigation of the effects of doping nitrogen atoms on superconductivity. Our simulations show that the lowest $T_c$ have been obtained at $LuH_3$ without doping nitrogen atoms. And the $T_c$ increased when doping concentration of N increasing. We find that doping N atoms to $LuH_3$ would increase its

$T_c$. But the highest $T_c$ in our calculation is 22 K with 1% N-doped concentration at 30 GPa, which is much lower than the level of room-temperature.

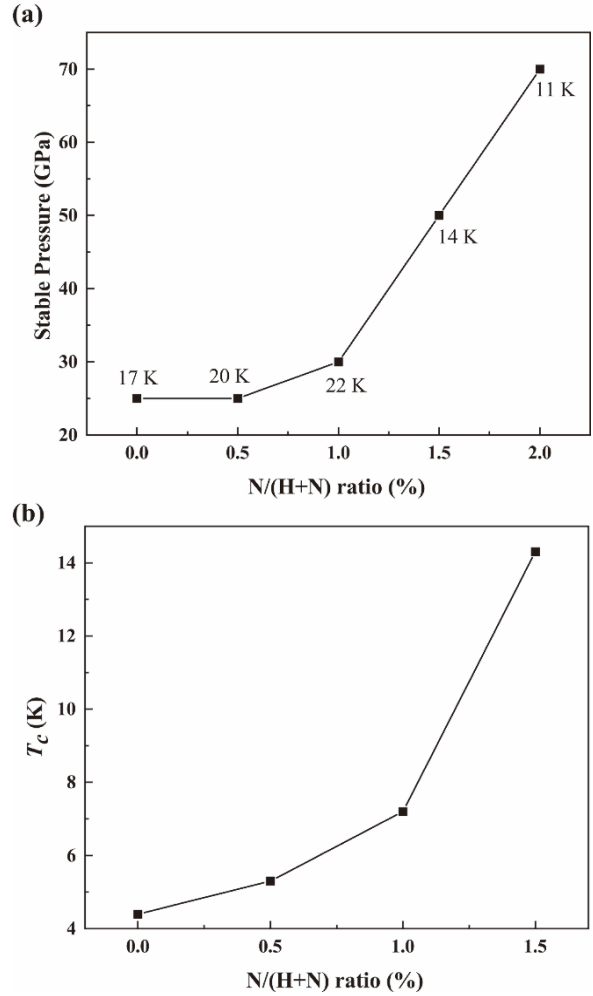

Fig. 4. (a) The minimum dynamically stable pressure dependence of the N-doped concentration. (b) The $T_c$ dependence of the N-doped concentration at 50 GPa. The Coulomb pseudopotential is using $\mu^*=0.10$.

Fig. 5(a-b) represented the band structure and density of states (DOS) of LuH$_3$ and LuH$_{2.97}$N$_{0.03}$ (1% N-doped concentration) at 50 GPa. The finite density of states at the Fermi level indicated the metallic feature of these structures. LuH$_3$ has a DOS near the Fermi level ($N(\epsilon_F)$) that reaches 0.269 states/eV/f.u. at its Fermi level. By substituting the H atoms with N, the LuH$_3$ system can be electron doped, which will raise the Fermi level and caused the bands near the G point to fall on top of the Fermi level. As a result, the Fermi level can be moved closer to the DOS peak and $N(\epsilon_F)$ would increase to 0.6

states/eV/f.u.. So introduction of N atoms can significantly enhance the metallic properties of LuH$_3$ by moving the Fermi level closer to the DOS peak.

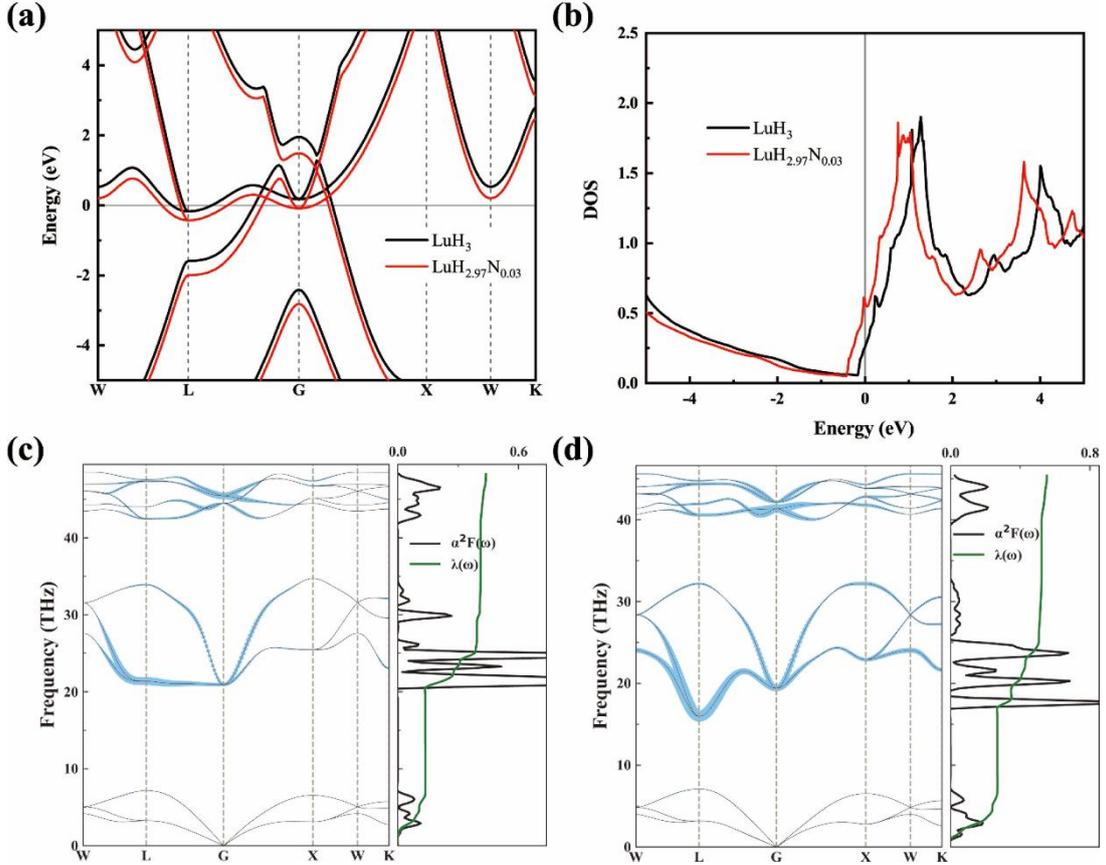

Fig. 5 (a-b) Electronic band structure and DOS (states/eV/f.u.) of LuH$_3$ (solid black) and LuH$_{2.97}$N$_{0.03}$ (solid red) at 50 GPa. The grey solid line means the Fermi energy. (c-d) Phonon spectra, Eliashberg phonon spectral function $\alpha^2F(\omega)$, and the electron-phonon integral $\lambda$ for LuH$_3$ and LuH$_{2.97}$N$_{0.03}$ at 50 GPa, respectively. The magnitude of the phonon linewidths is illustrated by the size of the blue circles.

Then we examine the phonon spectral and EPC of LuH$_3$ and LuH$_{2.97}$N$_{0.03}$ at 50 GPa (see Fig. 5(c-d)). The electron-phonon constant $\lambda$ is mainly contributed by the optical branch in LuH$_3$ (~0.301), while the acoustic branch has little contribution to it (~0.137). Doping with 1% nitrogen atoms can elevate $\lambda_{ac}$ (the interaction of EPC with acoustic phonons) from 0.137 to 0.268, while the $\lambda_{opt}$ (the interaction of EPC with optical phonons) remained almost unchanged. Therefore, the increase $T_c$ of N-doped LuH$_3$ is attributed to the enhancement of $\lambda_{ac}$. Interestingly, we find that the optical branch around L point in LuH$_3$ vary slightly with different wave vector $\boldsymbol{q}$ in the Brillouin zone.

And doping nitrogen atoms leads to a significant softening of this phonon mode. When the pressure decreases to 25 GPa (see Fig. S4 of Supplemental Material), the imaginary phonon frequency occurred near the L point indicate that the optical branch softening induced by N doping is responsible for the reduced stability of LuH$_3$.

## IV. Conclusions

In summary, we performed a first-principles study on the Lu-N-H system and find that there are no stable ternary compounds in this system at 1 GPa. Then we analysis the doping effect in N-doped $Fm\bar{3}m$-LuH$_3$ by using supercell method and VCA method. Even though the introduction of nitrogen atoms into $Fm\bar{3}m$-LuH$_3$ may slightly enhance its $T_c$, the introducing of N atoms would decrease its dynamics stability and their $T_c$ cannot reach the level of room-temperature. Furthermore, within the pressure range investigated in our study, the $T_c$ of N-doped $Fm\bar{3}m$-LuH$_3$ does not exceed 30 K. Therefore, our result shows that the N-doped $Fm\bar{3}m$-LuH$_3$ cannot be room-temperature superconductor.

## V. Acknowledgments

This work was supported by National Key R&D Program of China (No. 2018YFA0305900 and No. 2022YFA1402304), National Natural Science Foundation of China (Grants No. 12122405, 52072188 and 12274169), Program for Changjiang Scholars and Innovative Research Team in University (No. IRT_15R23), and Jilin Provincial Science and Technology Development Project (20210509038RQ). Some of the calculations were performed at the High Performance Computing Center of Jilin University and using TianHe-1(A) at the National Supercomputer Center in Tianjin.

## VI. Note added

We noted that during the preparation of our manuscript, the relevant theory research was reported by two different groups[36, 37].